\documentclass{ws-mpla}
\usepackage{url}
\usepackage[super]{cite}
\usepackage{graphicx}

\begin{document}

\markboth{Aldrin Cervantes and Miguel A. Garc\'ia-Aspeitia}
{Predicting cusps or kinks in Nambu-Goto dynamics}

\catchline{}{}{}{}{}

\title{Predicting cusps or kinks in Nambu-Goto dynamics}

\author{\footnotesize Aldrin Cervantes}
\address{Departamento de F\'isica, Cinvestav, Instituto Polit\'ecnico Nacional 2508,
San Pedro Zacatenco, 07360, Gustavo A. Madero, Ciudad de M\'exico, M\'exico\\
acervantes@fis.cinvestav.mx}

\author{Miguel A. Garc\'ia-Aspeitia}
\address{Consejo Nacional de Ciencia y Tecnolog\'ia, Av, Insurgentes Sur 1582. Colonia Cr\'edito Constructor, Del. Benito Ju\'arez C.P. 03940, M\'exico D.F. M\'exico.}
\address{Unidad Acad\'emica de F\'isica, Universidad Aut\'onoma de Zacatecas, Calzada Solidaridad esquina con Paseo a la Bufa S/N C.P. 98060, Zacatecas, M\'exico.\\
aspeitia@fisica.uaz.edu.mx}

\maketitle

\pub{Received (18 May 2015)}{Revised (Day July 2015)}

\begin{abstract}
It is known that Nambu-Goto extended objects present some pathological structures, such as cusps and kinks, during their evolution. In this paper, we propose a model through the generalized Raychaudhuri equation [Rh] for membranes to determine if there are cusps and kinks in the world-sheet. We extend the generalized Rh equation for membranes to allow the study of the effect of higher order curvature terms in the action on the issue of cusps and kinks, using it as a tool for determining when a Nambu-Goto string generates cusps or kinks in its evolution. Furthermore, we present three examples where we test graphically this approach.

\end{abstract}

\ccode{PACS Nos.: 04.20.Fy, 04.25.D-, 04.25.dc, 04.50.-h, 11.25.-w}
\section{Introduction}	
It is well known that when we consider a string described by the Nambu-Goto action it has some pathologies in its evolution, $i.e.$, cusps or kinks can be formed\cite{barba,greg}. In some cases the pathology of a closed string is widely used, for example, in topological defects with cosmic string these linear topological defects are predicted in a wide class of elementary particle models and could be formed as a symmetry breaking phase transition in the early universe\cite{vilen}. Also, these pathologies in general contribute to the emission of gravitational radiation\cite{vilenk}, in gravitational waves\cite{Thib} produced by cosmic strings, or as a cusp anomaly of a light-like Wilson line (see for example\cite{pol,brandt,craigie,korcherady,balitsky,korchem,korchemarche,bass,makee}).

Presently, we know that higher derivative models  involving the extrinsic curvature of the world-sheet help to remove pathologies in the evolution\cite{malc}, however there does not exist a method to know when such pathology arises in the Nambu-Goto dynamics. Here we present an approach to determine if a string will form cusps or kinks in its evolution. In turn, this model will help to study the evolution of an extended object to know when anomalies exist, as for example in brane cosmology\cite{branecos,Sahni,Maartens,Regge}.

In  previous works\cite{capoje}, the Raychaudhuri equation [Rh] was generalized to extremal relativistic membranes. In General Relativity [GR], where space-time curvature is related to matter, geodesic deviation provides a measure of the gravitational force, and the Rh equation plays an important role. One of its applications is to describe the focussing of  geodesics which is related to the singularities of space-time\cite{he}. For the case of membranes, the Rh equation is built on the world-sheet; thus, when we describe the Rh equation for membranes we are implicitly describing  the evolution of an extended object.

The Rh equation in relativistic membranes is useful in two different cases. The first case is when we have some background with a given singularity, then the Rh equation tells us how the evolution of the extended object is affected due to the singularity of the background\cite{kar}, and the second case is the one proposed here, where we use the Rh equation as a tool to determine the presence of such pathologies (cusps and kinks).

To this end, we extended the generalize Rh equation for extremal membranes to allow the study of the effect of higher order action on the issue of cusps and kinks and with the intention to make the model as general as possible. In this context, we use the Rh equation as a tool to determine if the world-sheet will exhibit cusps and kinks, whose Lagrangian depends on $L=L(g_{ab}\,,{K_{ab}}^i\,,\widetilde{\nabla}_a{{K}_{bc}}^i\,,\dots)$. The recipe consists in inserting the equations of motion obtained from the Lagrangian into the Rh equation. With this, we are able to analyze the behavior of \emph{extended Rh equation for membranes} and to relate the zeros of the solution with the cusps and kinks of the evolution of the string. Therefore, it is possible to predict whether the world-sheet will have cusps and kinks or not.

The paper is organized as follows. In Section \ref{geo}, we briefly review some geometric aspects of membranes; this section also introduces our notation. In Section \ref{gre}, We start giving a fairly complete overview of the formalism developed by Capovilla and Guven in \cite{capoje}; starting from Eq.~\eqref{geraeq}, extending the model to allow the study of the effect of higher order action. In Section \ref{ej} we show some examples which illustrate the use of the Rh equation to compute the existence of cusps and kinks, $i.e,$ points where the world-sheet will collapse forming cusps and kinks. Finally, in Section~\ref{conc} we present our conclusions and remarks, and we give a possible way to extend the model to relativistic $D$-dimensional membranes with $D>2$.
 
\section{Geometry formalism}\label{geo}

Consider an extended object $\Sigma$, of dimension $D$-$1$, evolving in a background space-time with $N$-dimension, with metric $\eta_{\mu\nu}$, $(\mu\,,\nu=0,1,\dots,N-1)$. The trajectory or world volume $m$, of the extended object is $D$-dimensional, and is described by the embedding:
\begin{equation}\label{parametri}
x^\mu=X^\mu(\xi^a)\,,
\end{equation}
where $x^\mu$ are local coordinates for the background space-time, $\xi^a$ are local coordinates for the world volume and $X^\mu$, are the embedding functions $(a\,,b=0,1,\dots,D-1)$. With the parameterization (\ref{parametri}) we obtain a basis of tangent vectors to the world volume at each point of $m$, denoted as:
\begin{equation*}
e_a=\frac{\partial}{\partial\xi^a}, \;\; e^{\mu}_a =X^{\mu}_{a} =\frac{\partial X^\mu}{\partial\xi^a}\,.
\end{equation*}
In this context, we introduce $N$-$D$, unitary normal vectors to the world volume denoted as ${n^\mu}_i$, $(i,j=1,2,\dots,N-D)$. Normal vectors to $m$ are implicitly defined by \footnote{Throughout this work, a midpoint denotes a contraction with the background metric $\eta_{\mu\nu}$. For example, $n^i\cdot e_a=\eta_{\mu\nu}n^{\mu i}\,{e^\nu}_a=0.$}, $n^i\cdot e_a=0$ and with normalization $n_i\cdot n_j=\delta_{ij}$. Thus, the vectors $\{e^{\mu}_{a},n^{\mu}_{i}\}$ form an orthonormal basis in the world volume $m$, adapted to $\Sigma$.\\

Throughout this paper we use latin letters for world volume and greek letters for the background. Notice also that all tensor in italics like $\mathcal{R}$, are also in the background and all tensor usually written like $R$ refer to the world volume. 

The metric induced (or first fundamental form) on the  world volume is then given by: $g_{ab}=e_a\cdot e_b=\eta_{\mu\nu}\,e^{\mu}_{a}\,e^{\nu}_{b}$.

Notice that we define the world sheet projections of the space-time covariant derivatives as $D_{a}:=e^{\mu}_{a} D_{\mu}$, where $D_\mu$ denotes the torsionless covariant derivative compatible with the space-time metric $\eta_{\mu\nu}$. Let us consider the world-sheet gradients of the basis vectors $\bf{\{e_a,n^i\}}$. Since they are space-time vectors, they are always decomposed using the orthonormal basis $\{e^{\mu}_{a}\,,n^{\mu}_{i}\}$\cite{spiva} as:
\begin{subequations}\label{gw}
\begin{equation}
D_a e_b=\gamma_{ab}^{c}\, e_c-K_{ab}^{i}\, n_{i}\,,
\end{equation}
\begin{equation}
D_{a} n^{i}=K_{ab}^{i}\,e^{b}+\omega_{a}^{ij}\,n_{j}\,,
\end{equation}
\end{subequations}
where $\gamma_{ab}^{c}$ is the world-sheet Ricci rotation coefficients and $K_{ab}^{i}$ is the $i$th extrinsic curvature of the world-sheet (or second fundamental form). Its symmetry in the tangential indices is a consequence of the integrability of the base $\{{e_a}\}$. Moreover, we have that $\omega_{a}^{ij}$ is the normal fundamental form (or twist potential) of the world-sheet.

The kinematical expressions in Eqs.~\eqref{gw}, which describe the extrinsic geometry, are generalizations of the classical Gauss-Weingarten equations, which altogether with the integrability conditions completely describe the extrinsic geometry of the world-sheet.

We therefore introduce a world-sheet covariant derivative defined on fields. It transforms as a tensor under normal frame rotations as
\begin{align}
\widetilde{\nabla}_{a}\Phi^{i_1\dots i_n}=&\nabla_a\Phi^{i_1\dots i_n}-\omega_{a}{}^{i_{1} j}\,\Phi_j{}^{i_2\dots i_n} - \cdots - \omega_{a}{}^{i_{n}j}\,\Phi^{i_1\dots i_{n-1}}{}_j\,,
\end{align}
where $\nabla_{a}$, is the intrinsic world-sheet covariant derivative. The gradients of the space-time basis $\{e_a\,,n^i\}$, along the directions orthogonal to the world-sheet, can be expressed as
\begin{subequations}
\begin{equation}
D_i n_j=\gamma_{ij}^{k}\, n_{k} - J_{aij}\,e^a\,,
\end{equation}
\begin{equation}
D_i e_a=J_{aij}\, n^j + S_{abi}\,e^b \,,
\end{equation}
\end{subequations}
where $D_{i}=n^{\mu}_{i}D_{\mu}$. The above relations are analogous to the Gauss-Weingarten Eqs.~\eqref{gw}. The quantities $J_{a}^{ij}$, and $S_{ab}^{i}$, are defined as:
\begin{align}\label{j}
J_{a}^{ij}&\equiv D^{i} e_{a}\cdot n^{j}\,,\\ \label{s}
S_{ab}^{i}&\equiv D^{i} e_{a}\cdot e_{b}=-S_{ba}^{i}\,,\\
\gamma_{_{ijk}}&\equiv D_i n_j\cdot n_k\,.
\end{align} 
In general, Eq.~\eqref{j} does not possess any symmetry under the interchange of the normal indices $i$, and $j$; this shows the fact that $\{n^i\}$ (unlike the $\{e_a\}$) do not generally form an integrable distribution. It is the analog of $K_{ab}^{i}$, in the Gauss-Weingarten equations. Moreover, Eq.~\eqref{s} is the analog of $\omega_{a}^{ij}$ and the Ricci rotation coefficient $\gamma_{ijk}$ is associated with the normal base.

\section{Generalization of Raychaudhuri equation}\label{gre}

For a geodesic curve, the Rh equation describes the evolution of $J^{ij}\equiv{J_0}^{ij}$, connecting neighboring geodesics along the curve and giving specific values to $J^{ij}$, at some initial time. The complete set of equations governing the evolution of deformations can be obtained by taking the gradient of $J_{ij}$, (see appendix of\cite{capoje} for an explicit calculation)
\begin{align}
\widetilde{\nabla}_b J_{a}^{ij}=&-\widetilde{\nabla}^{i}K_{ab}^{j}-J_{bk}^{i} J_{a}^{kj}-K_{bc}^{i} K_{a}^{cj} +\mathcal{R}_{\alpha\beta\mu\nu}\,n^{\alpha i}\,{e^\beta}_b\,{e^\mu}_a\,n^{\nu j}\,.\label{geraeq}
\end{align}
It should be noted that Eq.~\eqref{geraeq} does not depend on the equation of motion of the membrane. It is important to remark that in this point we differ from the original deduction done in\cite{capoje}, since they eliminate the term $\widetilde {\nabla} ^i K^j$, due to the fact that they only consider extremal membranes (\cite{capoje} only consider the extremal on the world-volume) whose equation of motion is $K^i=0$. From here, we will maintain the aforementioned term (do not necessarily vanishing) to allow the study of the higher order action on the issue of cusps or kinks. By taking the trace over the indices of the world-sheet we obtain
\begin{align}
\widetilde{\nabla}_a J^{aij}=&-\widetilde{\nabla}^i K^j- {{J_a}^i}_k J^{akj} - {K_{ac}}^i K^{acj} + \mathcal{R}_{\alpha\beta\mu\nu}\,n^{\alpha i}\,{e^\beta}_a\,e^{\mu a}\,n^{\nu j}\,.\label{najsi}
\end{align} 
If we now anti-symmetrize Eq.~\eqref{geraeq} with respect to its world-sheet indices, we get\begin{equation}\label{antisimej}
\widetilde{\nabla}_{a} J_{b}^{ij}-\widetilde{\nabla}_b J_{a}^{ij}=G_{ab}^{ij}\,,
\end{equation}
where we have defined
\begin{align*}
G_{ab}^{ij}\equiv& -J_{ak}^{i} J_{b}^{kj}-K_{ac}^{i} K_{b}^{cj} +\mathcal{R}_{\alpha\beta\mu\nu}\,n^{\alpha i}\,e^{\beta}_{a}\,e^{\mu}_{b}\,n^{\nu j}-(a\leftrightarrow b)\,.
\end{align*} 
It is worth noticing that the source term which includes $\widetilde{\nabla}^i K^j$, has been canceled out independently of the background dynamics.\\

Notice that it is not straightforward to work with the quantity $J_{a}^{ij}$, however, by analogy with mechanics of continuum, it is possible to decompose $J_{a}^{ij}$, into its symmetric and antisymmetric parts with respect to the normal indices $\Theta_{a}^{ij}$, and $\Lambda_{a}^{ij}$, respectively as
\begin{equation}\label{jtl}
J_{a}^{ij}=\Theta_{a}^{ij}+\Lambda_{a}^{ij}\,.
\end{equation}
We further decompose $\Theta_{a}^{ij}$, into its traceless and trace parts:
\begin{equation}\label{theta}
\Theta_{a}^{ij}=\Sigma_{a}^{ij}+\frac{1}{N-D}\delta^{ij}\Theta_{a}\,.
\end{equation}
In one dimension $\Theta$, $\Sigma^{ij}$, and $\Lambda^{ij}$, describe, respectively, the isotropic expansion, the shear and the vorticity of a trajectory with respect to neighboring trajectories. This is contained into the Rh equation in GR, but there is not a clear interpretation available for higher dimensions.

Now, we use Eq.~\eqref{jtl} and Eq.~\eqref{theta} into Eq.~\eqref{najsi}, in order to obtain the equation of motion for $\Theta_a$, $\Theta_{a}^{ij}$, and $\Lambda_{a}^{ij}$. Then it is possible to separate the symmetric and antisymmetric parts and the trace in the form
\begin{subequations}
\begin{align}
\widetilde{\nabla}_a\Sigma^{aij} + (\Lambda^{aik} \Lambda_{ak}^{j} + \Sigma^{aik}\Sigma_{ak}^{j})^{str} &+ \frac{2}{N-D} \Sigma_{a}^{ij}\Theta^a -[(M^2)^{ij}]^{str}=0\,,
\end{align}
\begin{align}
\widetilde{\nabla}_a\Lambda^{aij}+\Lambda^{ak[i}\Lambda_{ak}^{j]} &+ \Sigma^{ak[i}\Sigma_{ak}^{j]} - 2\Lambda^{ak[i}\Sigma_{ak}^{j]} + \frac{2}{N-D}\Lambda_{a}^{ij}\Theta^a=0\,,
\end{align}
\begin{align}
\widetilde{\nabla}_a\Theta^a - \Lambda^{aij}\Lambda_{aij} + \Sigma^{aij}\Sigma_{aij} &+ \frac{1}{N-D}\Theta_a\Theta^a - (M^2)^{i}_{i}=0\,,\label{the}
\end{align}
\end{subequations}
where the symbol $(\dots)^{str}$ denotes the symmetric traceless part of the matrix $(M^2)$ which is defined as:
\begin{align}
(M^2)^{ij}=-\widetilde{\nabla}^i K^j &-{K_{ab}}^i K^{abj} + \mathcal{R}_{\alpha\beta\mu\nu}\,n^{\alpha i}\,{e^\beta}_a\,e^{\mu a}\,n^{\nu j}.\,\label{matriz}
\end{align}
Now using Eq.~\eqref{jtl} and Eq.~\eqref{theta} into Eq.~\eqref{antisimej} and separating the symmetric and antisymmetric parts and the trace, we obtain:
\begin{subequations}
\begin{align}
2\widetilde{\nabla}_{[a}^{} \Sigma_{b]}^{ij} =&- 2(\Lambda_{[a}^{ik}\Lambda_{b]k}^{j} + \Sigma_{[a}^{ik} \Sigma_{b]k}^{j})^{str} + 4{\Lambda_{[a}}^{k(i}{\Lambda_{b]k}}^{j)}\,,
\end{align}
\begin{equation}
2\widetilde{\nabla}_{[a}^{} \Lambda_{b]}^{ij}= -2 \Lambda_{[a}^{k[i} \Lambda_{b]k}^{j]} - 2 \Sigma_{[a}^{k[i} \Sigma_{b]k}^{j]} - \Omega_{ab}^{ij}\,,
\end{equation}
\begin{equation}\label{isoexp}
2\partial_{[a}\Theta_{b]}=0\,.
\end{equation}
\end{subequations}
$\Omega_{ab}^{ij}$ is the curvature associated with the normal fundamental form $\omega_{a}^{ij}$, defined by,
\begin{equation*}
\Omega_{ab}^{ij}=\nabla_{b} \omega_{a}^{ij}-\nabla_{a} \omega_{b}^{ij}+\omega_{a}^{ik} \omega_{bk}^{j}-\omega_{b}^{ik} \omega_{ak}^{j}\,.
\end{equation*}
From Eq.~\eqref{isoexp}, which describes the evolution of the generalized expansion $\Theta_a$, it follows that: $\Theta_a=\partial_a\Upsilon$.
The above is implicit, at least locally, for some potential function $\Upsilon$. Inserting this expansion in Eq.~\eqref{the} we obtain
\begin{equation}\label{rayogex}
\Delta\Upsilon + \frac{1}{N-D}\partial_a\Upsilon\partial^a\Upsilon -\Lambda^2+\Sigma^2-M^2=0\,,
\end{equation}
where $\Delta$ is the Laplace-Beltrami operator defined as: $\Delta=\frac{1}{\sqrt{-g}}\partial_a(g^{ab}\sqrt{-g}\,\partial_b)$ and it is defined the world-sheet scalar quantities $\Lambda^2\equiv\Lambda^{aij}\Lambda_{aij}$, $\Sigma^2\equiv\Sigma^{aij}\Sigma_{aij}$ and $M^2\equiv{(M^2)^{i}}_{i}$, in Eq.~\eqref{rayogex} describes the evolution of the expansion of the world-sheet; having a quasilinear hyperbolic partial differential equation of second order. \\

It is possible to consider $\Upsilon$, as a generalized relative volume expansion potential. If $l$ represents the characteristic length of the expansion, we can set $\Upsilon=(N-D)\ln l$. With this change of variables, Eq.~\eqref{rayogex} becomes a linear equation
\begin{equation}\label{hipereq}
\Delta l+\frac{1}{N-D}\left(\Sigma^2-\Lambda^2-M^2\right)l=0\,.
\end{equation}
Now, Eq.~\eqref{hipereq} is the extended Rh equation for membranes, $i.e.$, it is a wave equation on the world-sheet for a massive positive definite scalar field $l$, with an effective mass term, $\mu^2=\left(\Sigma^2-\Lambda^2-M^2\right)/(N-D)$. Then, we have mapped the analysis of $\Theta_a$, to the solution of a linear wave equation. However, we should keep in mind that $\mu^2$ involves $\Sigma_{a}^{ij}$ and $\Lambda_{aij}$ explicitly and as a result, it will depend implicitly also on $\Theta_a$. We note that because  $M^2$ does not have a definite sign neither does $\mu^2$.\\

So far we have described the evolution of an extended object $\Sigma$, in a generic space-time background and we obtained the extended Rh equation for extended objects (see Eq. \eqref{hipereq}). We no longer have a particle moving in a space-time, now will be an extended object moving in a space-time. And we are interested in the geometry of that extended object.

\subsection{$M^2$ matrix analysis}

It is important to remark that Eq.~\eqref{hipereq} is the extended and generalized Rh equation for membranes, whose solution is not straightforward to find, due that it is necessary to determine $\Sigma^2,\,\Lambda^2$ and $(M^2)$, which depend of the dimension; however in some cases if the symmetric traceless part of $(M^2)^{ij}$ is zero, the following equation is fulfilled $\Sigma^2=\Lambda^2=0$. 
In particular, we analyze the form $(M^2)$ in Eq.~\eqref{matriz} and \eqref{hipereq} because in this paper we are interested on the \emph{evolution of the extended object}.

We start solving the second order hyperbolic partial differential equation \eqref{hipereq}, with the use of variable separation in order to obtain $D$-ordinary differential equations. Using the theorem on the existence of zeros\cite{tipler}, it is possible to obtain:
\begin{equation}\label{teo}
\frac{d^2z}{dt^2}+H(t)z=0\,,
\end{equation}
having at least one zero, if and only if $H(t)>0$. From this we can get an important physical information, to compare Eq.~\eqref{hipereq} with \eqref{teo} (after separation of variables). Notice that it is possible to see that both are second order differential equations with the possibility of identify: $\Sigma^2-\Lambda^2-M^2=H(t)$. One of the possibilities of obtain $\Sigma^2=\Lambda^2=0$, is if the symmetric traceless part of $(M^2)^{ij}$ is zero, concluding that the isotropic expansion $\Theta\to -\infty$, is possible if $M^2\leq 0$ in Eq.~\eqref{hipereq}. 

Then, we have reduced the problem of finding the conjugate points to the problem of discovering the location of zeros in solutions of Eq.~\eqref{hipereq} or Eq.~\eqref{rayhip} and just in the zeros of this solution is when we will have the collapse of the world-sheet in a region and therefore we have the cusps or kinks formation.\\

Continuing with our analysis of the trace of Eq.~\eqref{matriz} we have
\begin{equation}\label{matrizs}
(M^2)^{i}_{i}=-\widetilde{\nabla}^{i} K_{i}-K_{ab}^{i} K^{ab}_{i} + \mathcal{R}_{\alpha\beta\mu\nu}\,n^{\alpha i}\,e^{\beta}_{a}\,e^{\mu a}\,n^{\nu}_{i}\,.
\end{equation}
Using the integrability conditions of the Gauss-Codazzi\cite{capoje2}
\begin{equation}\label{gc}
\mathcal{R}_{\alpha\beta\mu\nu}\,e^{\alpha}_{a}\, e^{\beta}_{b}\, e^{\mu}_{c}\, e^{\nu}_{d} = R_{abcd} - K_{ac}^{i} K_{bdi} + K_{ad}^{i} K_{bci}\,,
\end{equation}
and substituting the completeness relationship $n^{\mu\,i} {n^\nu}_i=\eta^{\mu\nu}-e^{\mu a}\,{e^\nu}_a$, of the basis vector $\{{e^\mu}_a\,,{n^\mu}_i\}$ of the world volume $m$, in (\ref{matrizs}), and using (\ref{gc}) we find
\begin{align}\label{matrizf}
(M^2)^{i}_{i}=-\mathcal{R}_{\mu\nu}\,\mathcal{H}^{\mu\nu}-\widetilde{\nabla}^i K_i - 2 K_{ab}^{i} K^{ab}_{i} +K^{i}K_{i} + R\,.
\end{align}
We remark that this terminology is borrowed from the perturbative analysis of\cite{guve,cart,lars,capoje2}, where it appears as a variable mass in the world-sheet wave equation that describes the evolution of small perturbations and where $e^{a\mu}\,{e_a}^\nu=\mathcal{H}^{\mu\nu}$.

\subsection{Hypersurface}

In the case of a hypersurface, there is only one normal vector $i=1$, where $D=N-1$. In this case $G_{ab}^{ij}$, and $\omega_{a}^{ij}$, vanish identically due to the antisymmetry, since $\omega_{a}^{ij}=-\omega_{a}^{ji}$ and  $G_{ab}^{ij}=-G_{ab}^{ji}$ similar to $\Sigma^2$, and $\Lambda^2$; then, Eq.~\eqref{antisimej} becomes
\begin{equation}\label{hipersuj}
\partial_{a} J_{b}-\partial_{b} J_{a}=0\,,
\end{equation}
so that $J_{aij}:= J_{a11}\equiv\Theta_a$. Moreover, it is also possible to align the tangent vectors along the normal direction, so that $S_{ab}^{1}=0$, in an analogous way to a curve.

Therefore the Rh equation, in this special case, is reduced to,
\begin{equation}
\Delta\Upsilon + \partial_a\Upsilon\partial^a\Upsilon -M^2=0\,,
\end{equation}
again, setting $\Upsilon=(N-D)\ln l$, we have a hyperbolic second order partial differential equation,
\begin{equation}\label{rayhip}
\Delta l-\left(M^2\right)l=0\,.
\end{equation}
Then, the matrix $(M^2)$ is now modified. From Eq.~\eqref{matrizf}, the integrability conditions of Gauss-Codazzi\cite{capoje2} and the relation of completeness, $(M^2)$ is no longer a matrix
\begin{equation}\label{matrizh}
(M^2)=-\mathcal{R}_{\mu\nu}\,\mathcal{H}^{\mu\nu}- \nabla K - K^2 + 3R\,,
\end{equation}
where $\nabla$, is the covariant derivative. From Eq.~\eqref{matrizf} and Eq.~\eqref{matrizh}, we notice that $(M^2)$ depends on the background only in the first term, so if there is a singularity in the space background, this will  manifest itself in the membrane through the Ricci scalar, $\mathcal{R}$.

So far we have extended the Rh equation for extremal membranes for any codimension (see Eq. \eqref{hipereq}), and for the case of a hypersurface described by Eq.~\eqref{rayhip}, where the modification is through the matrix $(M^2)$, since in the original deduction made by Capovilla and Guven, they only considered extremal membranes; therefore they assume $K^i=0$, from Eq. (\ref{geraeq}). 
However, if we want to study the effect of other action (as in the example of Sec. \ref {clng}) or a higher order action, we need that the term $K^i\neq0$ is fulfilled, even more we have that $\widetilde{\nabla}^i K_i\neq0$, in a more general way.

With the aim of illustrating the method, we address three examples for determining the presence of cusps or kinks in the world-sheet.

\section{Testing the Model: Examples}\label{ej}

The evolution of the string is a two-dimensional extended object (world-sheet), in this context the extended Rh equation for membranes tells us, the singularities on the world-sheet (if exist) $i.e.,$ the string collapses, and with this new method we can predict if the extended object could have cusps or kinks during its evolution. In order to determine this, we take the following steps:

\begin{itemize}
\item We start from the action and make the variation, then we introduce the equation of motion in Rh's equation, Eq.~\eqref{hipereq} (or in Eq.~\eqref{rayhip} for the case of the hypersurface).

\item Given the parametrization of the string, we obtain the corresponding geometric tensors that describe the world sheet.

\item Substituting the parameterization in the equation of motion from the action, we get the solution and replace it back into the Rh equation.

\item We solve the differential equation through the method of variables separation, making $l(\tau,\sigma)=\ell(\tau)\varphi(\sigma)$; the temporal part of the solutions is analyzed by checking if the solution cross or not the $\tau$-axis. 

\item We will have two cases: a) crossing $\tau$-axis, in this case the world sheet collapses, implying the existences of cusps or kinks. b) not crossing $\tau$-axis, in this case the world sheet never collapses showing the stability of the structure.
\end{itemize}

In the following, we study the \emph{helical string in breathing mode} and the model of \emph{circular loop} to observe the stability (or not) of the world sheet and identify the existence (or not) of the cusps or kinks.

\subsection{Helical string in breathing mode}

We will take the following simple case: consider extreme membranes, which satisfy the Nambu dynamics\cite{nambu}. The Nambu action is proportional to the area of the world sheet:
\begin{equation}\label{dng}
S[X^\mu]=-\alpha\int_m d^D\xi\,\sqrt{-g}\,,
\end{equation}
where $\alpha$ is the tension of the membrane and its equation of motion is given by
\begin{equation}\label{ecmovnambu}
K^i\equiv g^{ab}K_{ab}^{i}=0\,.
\end{equation}
Substituting $K^{i}=0$ in Eq.~\eqref{matrizf}, $(M^2)$ takes the form $(M^2)=-\mathcal{R}_{\mu\nu}\,\mathcal{H}^{\mu\nu}+R$; therefore, the Rh equation (\ref{hipereq}) for this case is
\begin{equation}\label{eqrheli}
\Delta l+\frac{1}{N-D}(\mathcal{R}_{\mu\nu}\,\mathcal{H}^{\mu\nu}-R)\,l=0\,.
\end{equation}
Now we consider the \emph{helical string in breathing mode}, $\Sigma$, evolving into a Minkowski background space $(3+1)$ dimensional, parameterized as follows\cite{maljoh}
\begin{equation}\label{parahelical}
x^\mu=X^\mu(\xi^a)=(\tau\,,Z(\tau)\,\cos\sigma\,,Z(\tau)\,\sin\sigma,q\,\sigma)\,,
\end{equation}
where the tangent vectors will be given by $e^{\mu}_{a}\, (a=\tau,\sigma)$:\begin{align*}
&e^{\mu}_{\tau} =\left(1\,,\dot{Z}\,\cos\sigma\,,\dot{Z}\,\sin\sigma,0\right)\,,\\
&e^{\mu}_{\sigma} =(0\,,-Z\,\sin\sigma\,,Z\,\cos\sigma,q)\,,
\end{align*}
the dot denotes differentiation with respect to $\tau$, This string is helical with breathing $q$, and notice that in the limit $q\to 1$, corresponds to the flat world-sheet and the limit $q\to 0$, to the collapsing circular loop.

The induced metric $g_{ab}=\eta_{\mu\nu}e^{\mu}_{a}e^{\nu}_b$ will be now given by,
\begin{equation}
ds_\Sigma^2=g_{ab}\,d\xi^a d\xi^b=-(1-\dot{Z}^2)d\tau^2+(Z^2+q^2) d\sigma^2.
\end{equation}
On the other hand, the normal vectors are:
\begin{align*}
&n^{\mu}_{1}=\frac{1}{\sqrt{q^2+Z^2}}\left(0\,,q\,\sin\sigma\,,-q\,\cos\sigma,Z\right)\,,\\
&n^{\mu}_{2}=\frac{1}{\sqrt{1-\dot{Z}^2}}\left(\dot{Z}\,,\cos\sigma\,,\sin\sigma,0\right).
\end{align*}
In addition, the extrinsic curvature will be
\begin{center}
$K_{1ab}=\displaystyle\frac{q\,\dot{Z}}{\sqrt{q^2+Z^2}}\begin{pmatrix}
             0 & 1 \\
             1 & 0
\end{pmatrix}$\,,\qquad 
$K_{2ab}=\displaystyle\frac{1}{\sqrt{1-\dot{Z}^2}}\begin{pmatrix}
             -\ddot{Z} & 0 \\
                        0 & Z
\end{pmatrix}$\,,
\end{center}
and the normal fundamental form of the world-sheet is given by
\begin{center}
$\left(\omega_{a}^{ij}\right) = \displaystyle \frac{q}{\sqrt{(q^2+Z^2)(1-\dot{Z}^2)}}\begin{pmatrix}
             0 \\
             1
\end{pmatrix}$\,.
\end{center}
Using the parameterization showed in Eq.~\eqref{parahelical} on the equation of motion described by Eq.~\eqref{ecmovnambu}, we find the following differential equation
\begin{equation*}
\frac{\ddot{Z}}{{(1-\dot{Z}^2)}^{3/2}}+\frac{Z}{(q^2+Z^2)(1-\dot{Z}^2)^{1/2}}=0\,,
\end{equation*}
whose general solution can be written as
\begin{equation*}
Z(\tau)=\sqrt{\kappa^2-q^2}\,\cos\left(\frac{\tau-\tau_0}{\kappa}\right).
\end{equation*}
With the aim to illustrate the method we will reduce some constants, and therefore choosing the initial condition $\tau_0=0$ and $\kappa=1$, then we substitute the solution $Z(\tau)$ in Eq.~\eqref{eqrheli} and takes the form,
\begin{align*}
\frac{\partial^2}{\partial\sigma^2}l(\tau,\sigma)&-\frac{\partial^2}{\partial\tau^2}l(\tau,\sigma) +\left[\frac{(q^2-1)\,(-\cos^2\tau+q^2\sin^2\tau)}{\left(\cos^2\tau+q^2\sin^2\tau\right)^2}\right]\,l(\tau,\sigma)=0\,.
\end{align*}

Notice that the term $\partial^2l(\tau,\sigma)/ \partial\tau\sigma$, does not appear due to the fact that the metric is diagonal ($g_{\tau\sigma}=0$). Now, we use separation of variables $l(\tau,\sigma)=\ell(\tau)\varphi(\sigma)$ to obtain the equation for $\tau$ and $\sigma$. The equation for $\sigma$ is 
\begin{equation*}
\frac{1}{\varphi(\sigma)}\frac{d^2}{d\sigma^2}\varphi(\sigma)=m^2\,,
\end{equation*}
that is an ordinary differential equation (ODE's) whose solutions is $\varphi(\sigma)=e^{\pm m\sigma}$. Now we must distinguish that we have three cases for the constant of separation $m$: One for $m=0$, other for $m<0$, and the other one for $m>0$. After a numerical analysis of the world-sheet, we found that only when $m>0$, the isotropic expansion $\Theta\to -\infty$, produce the scenario that we are looking for (i.e. the geodesic congruence).

Since we are interested in the evolution of the system, we focus on $\ell(\tau)$, because $\tau$, is the proper time of the string:
\begin{equation}\label{solhsbm}
\frac{d^2}{d\tau^2}\ell(\tau)-\left[\frac{(q^2-1)\,(-\cos^2\tau+q^2\sin^2\tau)}{\left(\cos^2\tau+q^2\sin^2\tau\right)^2}+m^2\right]\,\ell(\tau)=0\,,
\end{equation}
where $m^2$, is the constant of separation, for intermediate $q$, the trajectory is never singular, and the extrinsic curvature peaks at approximately $q^{-2}\sqrt{1-q^2}$. Eq.~\eqref{solhsbm} reminds us the equation of an oscillator, but with the difference that the term in square brackets explicitly depends on $\tau$, into the oscillator problem this term is a constant and therefore, our solution is no longer oscillatory. However, Eq.~\eqref{solhsbm} look like Eq.~\eqref{teo} with the term in brackets equal to $H(t)$, then we can apply the theorem on the existence of zeros of ordinary differential equations.

\begin{figure}[!h]
\centerline{\includegraphics[width=11.5cm]{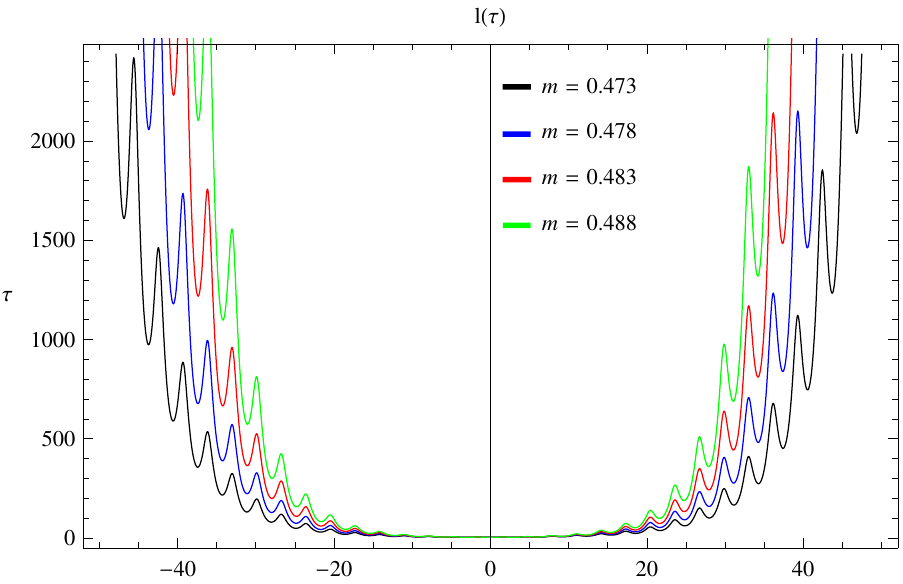}}
\centerline{\includegraphics[width=11.5cm]{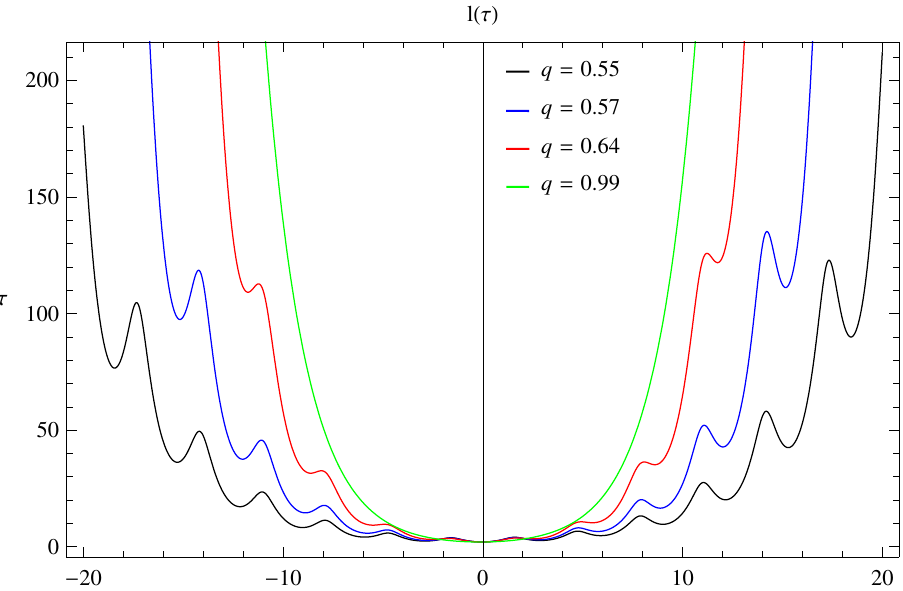}}
\caption{Numerical solution of Eq.~\eqref{solhsbm}, with initial condition $\tau_{_0}=0$ for $\kappa=1$ and $q = 0\text{.}55$, where it is observed the breathing frequency of the helix. (Top) In this case we vary little bit the values of $m$, due to this increase quickly. (Bottom) Here the parameter $0\le q <1$ determines its winding number per unit length, when $q\to 0$ oscillations are greatest and where $q\to 1$ oscillations tend to disappear. Both figures show the non existence of cusps and kinks. From the plot we see that the helical breathe is a time-dependent solution which is never singular. See the text for more details. \protect\label{fig:1}}
\end{figure}

In Fig.~\ref{fig:1}, one can see the Rh equation of the helical string in breathing mode, in Minkowski space-time background. In the Top panel we plot the solution with different values of the constant of separation $m$, while in the Bottom panel we plot different values of $q$, where $q$ is the winding number per unit length. The oscillations of the graphic are due to breathing mode (and this is precisely what prevents the collapse). Notice that the graph never crosses the $\tau$-axis, which implies that the world-sheet will never have a cusps or kinks in the evolution, which is in agreement with\cite{sake}.

\subsection{Circular loop}\label{cp}
 
In this example, we start writing down the Nambu-Goto action (see Eq.~\eqref{dng}), whose equation of motion is given by Eq.~\eqref{ecmovnambu}. Here, we consider the hypersurface of a circular loop, $\Sigma$, evolving into a $(2+1)-$dimensional Minkowski space-time background, which is parameterized as follows\cite{maljoh}

\begin{equation}\label{pacirc}
x^\mu=X^\mu(\xi^a)=(\tau\,,Z(\tau)\,\cos\sigma\,,Z(\tau)\,\sin\sigma)\,,
\end{equation}
where the tangent vectors are now given by $e_{a}^{\mu}\,(a=\tau,\sigma)$,
\begin{align*}
&e^{\mu}_{\tau} =(1\,,\dot{Z}\,\cos\sigma\,,\dot{Z}\,\sin\sigma)\,, \\
&e^{\mu}_{\sigma} =(0\,,-Z\,\sin\sigma\,,Z\,\cos\sigma)\,.
\end{align*}
In this particular case, the induced metric $g_{ab}=\eta_{\mu\nu}e^{\mu}_{a}e^{\nu}_{b}$ will be
\begin{equation}
ds_\Sigma^2=g_{ab}\,d\xi^a d\xi^b=-(1-\dot{Z}^2)d\tau^2+Z^2 d\sigma^2\,.
\end{equation}
and the normal vector takes the form:
\begin{equation*}
n^\mu=\frac{1}{\sqrt{1-\dot{Z}^2}}\left( \dot{Z}, \cos\sigma, \sin\sigma \right)\,,
\end{equation*}
then, the extrinsic curvature will be
\begin{equation*}
K_{\tau\tau}=-\frac{\ddot{Z}}{\sqrt{1-\dot{Z}^2}}, \;\; K_{\sigma\sigma}=\frac{Z}{\sqrt{1-\dot{Z}^2}}\,.
\end{equation*}
Replacing the embedding relation written in Eq.~\eqref{pacirc} into Eq.~\eqref{ecmovnambu}, it takes the specific form
\begin{equation}
\ddot{Z} Z-\dot{Z}^2+1=0\,,
\end{equation}
whose general solution is:
\begin{equation}
Z(\tau)=\kappa\,\cos\left(\frac{\tau-\tau_0}{\kappa}\right)\,.
\end{equation}
In order to solve the differential equation we have chosen $\tau_0=0$, $\kappa=1$ as initial condition, thus one obtains the canonical form of the loop trajectory \eqref{pacirc}, substituting $K=0$ in the equation of the hypersurface, Eq.~\eqref{matrizh}, and finding that $(M^2)$ takes the form $(M^2)=3R$; therefore, the Rh equation (\ref{rayhip}) for this case is
\begin{equation*}
\Delta l-(3R)\,l=0\,.
\end{equation*}
Substituting the solution $Z(\tau)$ in the above equation we obtain
\begin{equation*}
\frac{\partial^2}{\partial\sigma^2}l(\tau,\sigma)-\frac{\partial^2}{\partial\tau^2}l(\tau,\sigma)+6\,\sec^2\tau\,l(\tau,\sigma)=0\,,
\end{equation*}
again, using separation of variables $l(\tau,\sigma)=\ell(\tau)\varphi(\sigma)$, as above, the equation for $\varphi(\sigma)$ is ODE's, and where $m^2$ is the new constant of separation.

Then the equation for $\ell(\tau)$, becomes
\begin{equation}\label{soldngcirc}
\frac{d^2}{d\tau^2}\ell(\tau)-\left(6\,\sec^2\tau-m^2\right)\ell(\tau)=0\,,
\end{equation}

\begin{figure}[!h]
\centerline{\includegraphics[width=11.5cm]{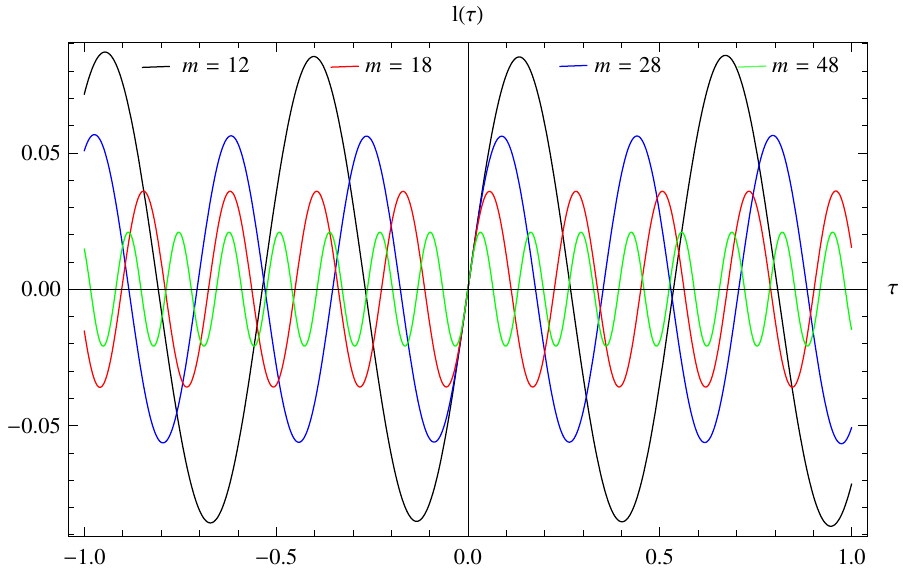}}
\caption{Numerical solution of Eq.~\eqref{soldngcirc}, with initial condition $\tau_{_0}=0$ for $\kappa=1$ and $\Sigma^2=\Lambda^2=0$. Were $m$ is proportional to frequency and inversely proportional to the amplitude. In this case, it is shown the existence of cusps/kinks due that graph passes on the $\tau$-axis. See the text for more details.\protect\label{fig:3}}
\end{figure}

We solve Eq.~\eqref{soldngcirc} numerically and plot the solution in Fig.~\ref{fig:3}. We see that the \emph{circular loop} form cusps and kinks in the world-sheet, because of the graph cuts the $\tau$-axis and, in this way, the circular loop collapses at a point. The extrinsic curvature invariants, become singular at this point, hence rigidity would be indicated by a retardation of the collapse or a positive correction to the amplitude of the loop. The solution that we find is also in a good agreement with\cite{barba}. 

\subsection{Circular loop in the first NG correction}\label{clng}

With the intention to show that the model is applicable in different situations, we will consider the first correction to NG action written as
\begin{equation}\label{ack}
S[X]=\alpha \int_m d^D\xi\,\sqrt{-g}\,K\,,
\end{equation}
where $\alpha$ is the tension of the membrane, and its equation of motion is given by
\begin{equation}\label{solpng}
R=0\,.
\end{equation}
whose equation of motion remains second order. Again we consider the hypersurface of a circular loop, $\Sigma$, evolving into a $(2+1)-$dimensional Minkowski space-time background, parameterized as in section \ref{cp}. Substituting $R=0$ in the equation of the hypersurface Eq.~\eqref{matrizh}, $(M^2)$ takes the form $(M^2)=-\nabla K - K^2$, where $\nabla$, is the covariant derivative; therefore, the Rh equation (\ref{hipereq}) for this case is
\begin{equation}\label{rheqfng}
\Delta l+(\nabla K + K^2)\,l=0\,.
\end{equation}

\begin{figure}[!h]
\centerline{\includegraphics[width=11.5cm]{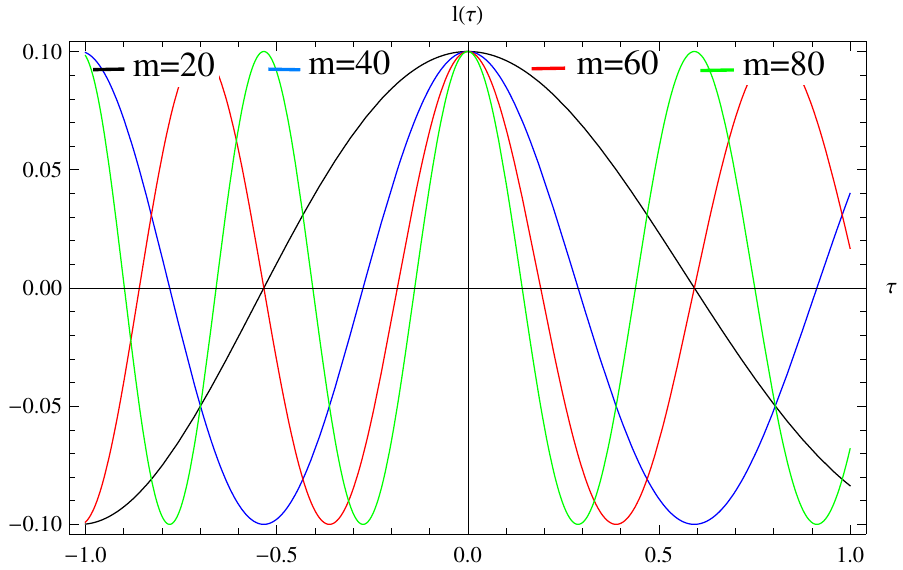}}
\caption{Numerical solution of Eq.~\eqref{solfngcirc}, with the condition $\Sigma^2=\Lambda^2=0$. Where $m$ is proportional to frequency;  $a$ and $b$ are inversely proportional to frequency. In this case, it is shown the existence of cusps/kinks due that graph passes on the $\tau$-axis.}\protect\label{fig:5}
\end{figure}

Replacing the embedding relation written in Eq.~\eqref{pacirc} into Eq.~\eqref{solpng}, it takes the specific form
\begin{equation}
\frac{2\,\ddot{Z}}{Z\,(\dot{Z}^2-1)^2}=0\,,
\end{equation}
whose general solution is:
\begin{equation*}
Z(\tau)=a+b\,\tau\,,
\end{equation*}
where $a$ and $b$ are integration constants. Substituting the solution $Z(\tau)$ in Eq.~(\ref{rheqfng}) we obtain
\begin{align*}
\frac{\partial^2}{\partial\sigma^2}l(\tau,\sigma) + \frac{b\,\sqrt{1-b^2}-1}{b^2-1}\,l(\tau,\sigma) + \frac{b(a+b\tau)}{b^2-1}\frac{\partial}{\partial\tau}\,l(\tau,\sigma) + \frac{(a+b\tau)^2}{b^2-1}\frac{\partial^2}{\partial\tau^2}\,l(\tau,\sigma)=0.
\end{align*}
Using the method of separation of variables as above, we have again two equations, one for $\ell(\tau)$ and the other for $\varphi(\sigma)$, thats its a ODE's whit $m^2$ the new constant of separation. Focusing on the evolution equation, we have for $\ell(\tau)$ the following equation:
\begin{equation}\label{solfngcirc}
\frac{d^2}{d\tau^2}\ell(\tau)+\frac{b}{a+b\tau}\frac{d}{d\tau}\ell(\tau)+\frac{m^2(b^2-1)}{(a+b\tau)^2}\,\ell(\tau)=0\,.
\end{equation}
We solve Eq.~\eqref{solfngcirc} numerically and plot the solution in Fig.~\ref{fig:5}. We see that the \emph{circular loop} form cusps and kinks in the world-sheet, because of the graph cuts the $\tau$-axis and, in this way, the circular loop collapses at a point.

It is remarkable that the circular loop continues collapsing but the frequency of the collapse is different, is for it that when we consider more orders in the action the collapse runs but does not disappear, it happens up to fourth order in the action\cite{malc}.

\section{Conclusions and Remarks}\label{conc}

As a first step we have provided a new model to determine if a string will present or not cusps and kinks, we do this by examining the evolution of deformations of relativistic world-sheet, propagating in a background space-time of arbitrary codimension. The construction of the model was motivated by the Rh equation for relativistic membranes, although the method is algebraic, so a numerical analysis is needed to determine the solution of the final differential equation.

In GR, the most general way to determine if a space-time is singular is by the incompleteness geodesic condition which is determined by using the Rh equation, and by analyzing the global structure of the space-time. In relativistic membranes, this condition is given by Eq.~\eqref{hipereq} or Eq.~(\ref{rayhip}) which have a general structure. It is important to remark that they contain information about the background, across the Ricci tensor and also possess information about the membrane via the extrinsic curvature and the scalar curvature. With these equations it is possible to determine if the world sheet collapse $i.e.$, there are formation of cusps or kinks.

In addition we extended the generalize Rh equation from membranes, where the modification is through the matrix $(M^2)$. It is important to note that in the original deduction made by Capovilla and Guven, they only consider extremal membranes whose equation of motion is $K^i=0$. Notice that if we want to study the effect of other geometries or higher order action, the term $K^i$ is present and the most natural extension of the model is including this situations as in the case of example \ref{clng}.

In the example presented in section \ref{cp}, we could appreciate the importance of the Rh equation for membranes. Since we do not necessarily have a background that has some singularity for what the extended object collapses. In fact we did not need to fix the background nor even study the singularities of it, some times the evolution of the extended object is unstable (form cusps or kinks) regardless of the background. This means that the extended object may have singularities (cusps and kinks) in the development by itself, which is a pathology that exhibits Nambu-Goto action, and it would be interesting to investigate in subsequent works, if this behavior is inherent in more dimensions.

With all this, we would have a general method for determining whether a system based on the extrinsic curvature model will have cusps or kinks in its evolution. 

Finally as a second step, one of our future interest is to apply this method to brane cosmology. Where we start with Nambu-Goto action but in more dimensions and will make corrections as we did in Sec. \ref{clng}. However, this is ongoing research that will be presented elsewhere. 
\section*{Acknowledgments}
AC thanks Julio Hern\'andez and Jukka Sarvas for help and discussions on numerical analysis. AC also thanks Laura G\'omen, Riccardo Capovilla and Efra\'in Rojas for helpful discussion. AC acknowledges support from CONACyT scholarship (M\'exico). MAG-A acknowledges support from C\'atedra-CONACYT and SNI. We also thank the support of Instituto Avanzado de Cosmolog\'ia (IAC) and the Beyond Standard Theory Group (BeST).


\end{document}